# Quantum-Secure Physical Unclonable Function enabled by Silicon Photonics Integrated Circuits

G. Sarantoglou, N. Tzekas, G. Moustakas, G.A. Karydis, V. Kaminski, E. Protsenko, K. Gradkowski, A. Bazin, C. Vigliar, A. Bogris and C. Mesaritakis

***Abstract*— Physical Unclonable Functions (PUFs) are hardware security primitives whose inherent physical complexity can be exploited for secure authentication and cryptographic key generation. Silicon photonic devices, owing to their suitability for quantum and artificial intelligence applications alongside standard CMOS fabrication processes, constitute a highly promising substrate for integrated multifunctional PUFs. Despite the advanced security guarantees offered by quantum cryptographic protocols and the central role of silicon photonics in quantum technologies, quantum readout strategies based on single-photon states for photonic PUFs remain largely unexplored. In this work, we experimentally demonstrate a silicon nitride (SiN) programmable photonic Mach–Zehnder interferometer mesh that implements a unitary transformation and operates as a PUF, whose secret physical signature arises from uncontrollable waveguide variations during fabrication. Using experimentally derived parameters from the SiN integrated mesh, we further introduce and numerically evaluate a quantum readout protocol that combines single-photon states with PUFs. Maximally mixed quantum states are employed to conceal the underlying unitary transformation from passive eavesdropping. Security against adversaries possessing devices fabricated under similar conditions is assessed, with authentication performance quantified through Monte Carlo analysis of the false acceptance and false rejection rates as a function of the number of detected events and corrected errors. The results indicate exceptional performance with equal error rates as low as $10^{-14}$, highlighting the potential of quantum-secure PUFs for high-security authentication applications.**

## I. INTRODUCTION

PHYSICAL unclonable functions (PUF) have received increased attention in the ever-growing landscape of densely interconnected devices, promising strong hardware-rooted security for systems with limited computational and memory resources [1], [2]. In this context, authentication is a fundamental procedure for establishing a secure communication channel between the user and a trusted entity relying on the possession of a unique physical token. In conventional authentication protocols, a unique digital secret key is generated through algorithmic driven Pseudo-Random Number Generators (PRNGs) and stored in a non-volatile memory. This key is presented to the authenticator whenever communication needs to be established. Two critical vulnerabilities of these approaches are either statistical anomalies in key generation through the use of hardware restrained PRNGs [3] or key theft through tampering attacks at the non-volatile media [4]. PUFs address this weakness by shifting the security primitive from the digital to the physical domain. Specifically, PUFs exploit the intrinsic and uncontrollable physical variations of a device, by applying to it a set of analog input signals provided

This work was supported in part by PROMETHEUS (ID: 101070195), QPIC1550 (ID: 101135785) Horizon Europe projects. G. Sarantoglou is supported by the Project QUASAR which is implemented in the framework of H.F.R.I call “Basic research Financing (Horizontal support of all Sciences)” under the National Recovery and Resilience Plan “Greece 2.0” funded by the EU – Next Generation EU No: 016594. Corresponding author: G. Sarantoglou.
G. Sarantoglou is with the Department of Biomedical Engineering, University of West Attica, Agiou Spiridonos,12243, Egaleo, Greece and Dept. Information and Communication Systems Engineering, University of the Aegean, Palama 2 str. Karlovasi Samos 83200-Greece (e-mail: gsarantoglou@uniwa.gr). N. Tzekas and C.Mesaritakis are with Department of Biomedical Engineering, University of West Attica, Agiou Spiridonos,12243, Egaleo, Greece (email: ntzekas@uniwa.gr, cmesar@uniwa.gr ). G.Moustakas, G.A. Karydis and A. Bogris are with Department of Informatics & Computer Engineering, University of West Attica, Agiou Spiridonos 12243, Egaleo, Athens, Greece (email: gemoustakas@uniwa.gr, gakarydis@uniwa.gr, abogris@uniwa.gr ). Veronika Kaminski, Ekaterina Protsenko and Caterina Vigliar are with Department of Electrical and Photonics Engineering, Technical University of Denmark, DK 2800 Kgs. Lyngby, Denmark (email: veronik@dtu.dk, ekprot@dtu.dk, catvi@dtu.dk ) .Kamil Gradkowski is with the Tyndall National Institute, Lee Maltings Complex, Dyke Parade, T12 R5CP, Cork, Ireland (email: kamil.gradkowski@tyndall.ie). Alexander Bazin is with LIGENTEC France, 224 Boulevard Kennedy, 91100 Corbeil-Essonnes, France (email: alexander.bazin@ligentec.com ).

by the authenticator, known as challenges, and processing on demand the analog outputs to generate the response, that is the authentication key [2]. Since the key is not digitally stored in the memory but it is instead generated on demand from the device's physical structure, PUFs provide enhanced protection against key extraction and eavesdropping attacks.

The most crucial vulnerability of classical PUFs is the malevolent collection of challenge-response pairs (CRPs), that permits the emulation of the device's functionality [2]. The fundamental root of this challenge is the cloning of classical states via a side channel attack, that is a shared vulnerability both for electronic [5], [6], [7] and photonic PUFs [8], [9], [10]. Quantum security algorithms, under-pinned by the no-cloning theorem [11] and the fundamental non-distinguishability of non-orthogonal quantum states provide powerful security primitives for modern cryptographic protocols, as it can be witnessed by the ever-growing advancement of quantum key distribution algorithms (QKD) [12]. However, such mechanisms have yet to be successfully transferred to the PUF context.

Typical electronic PUFs are not easily adapted to a quantum regime of operation. Quantum PUFs, whose fingerprint is linked to the gate oxide thickness of standard MOSFET transistors and the corresponding quantum tunnelling current, are interrogated in a "classical" manner thus not offering the non-cloning property [13]. On the other hand, purely fermionic PUFs are bulky and require extreme cooling schemes [14]. Quantum optics provides an alternative approach, leveraging CMOS-compatible silicon photonics and the inherent stability of photon states that enables practical, room-temperature quantum computing [15]. Quantum security has been extended to optical PUFs using non-classical light sources and single-photon detectors, giving rise to quantum CRPs [16]. In such schemes, any measurement performed by an eavesdropper unavoidably disturbs the quantum channel, rendering interception attempts detectable and hinders the simultaneous monitoring of both the challenge and the response that fuels emulation-based attacks. Quantum-secure PUFs have been presented experimentally only for free-space optical setups, which severely limits their scalability and practical deployment [17], [18], [19]. Despite the rapid adoption of silicon photonics in quantum computing industry, Quantum-secured Silicon Photonic PUFs (QSP-PUFs) have yet to be systematically analyzed, with their merits remaining largely unexplored [20]. In this work, we present the case of a silicon nitride photonic programmable mesh as a SP-PUF [21] that implements a unitary transformation and equip it with quantum security properties by introducing single photon sources and detectors. In this case, the security key remains in the fabrication variations of the waveguides, yet the scheme is supported by two fundamental quantum concepts; the maximally mixed state (MMS) that conceals CRPs for any party other than a trusted verifier and the no-cloning theorem that inhibits copying the state without measurement. The manuscript is organized as follows: In the first section, the programmable mesh is experimentally verified as a PUF. The photonic integrated circuit (PIC) is fabricated by LIGENTEC using the AN800multi-project wafer run (200 mm wafer production line). The waveguide cross section of 1.0 x 0.8 $\mu m$ (width x height) constituted the core guiding layer and is coupled to a top metal line used for thermo-optical phase control. This device presents excellent randomness with a fractional inter-Hamming distance of 45 % and robustness with a fractional intra-Hamming distance of 2 %. In the second section, the QSP-PUF is presented under two attack scenarios: (a) side channel attack based on passive eavesdropping targetting the secret fingerprint and (b) possession of a PIC fabricated on the same wafer as the legitimate device, assessing the level of trust in the manufacturer. The first attack is resolved by the MMS and the no-cloning theorem that conceal the fingerprint of the legitimate device. For the second attack, we evaluate the probability of cloning using the experimentally derived parameters from the first section, yielding the false acceptance rate (FAR). We further assess the robustness of the legitimate device to obtain the false rejection rate (FRR). Both error rates are computed as functions of the number of detected clicks and the corrected errors, which act as hyperparameters tuned by the verifier. Equal FAR and FRR values as low as $10^{-14}$ are achieved, demonstrating that even PICs fabricated from the same wafer exhibit substantial physical variability, while maintaining excellent robustness against environmental noise.

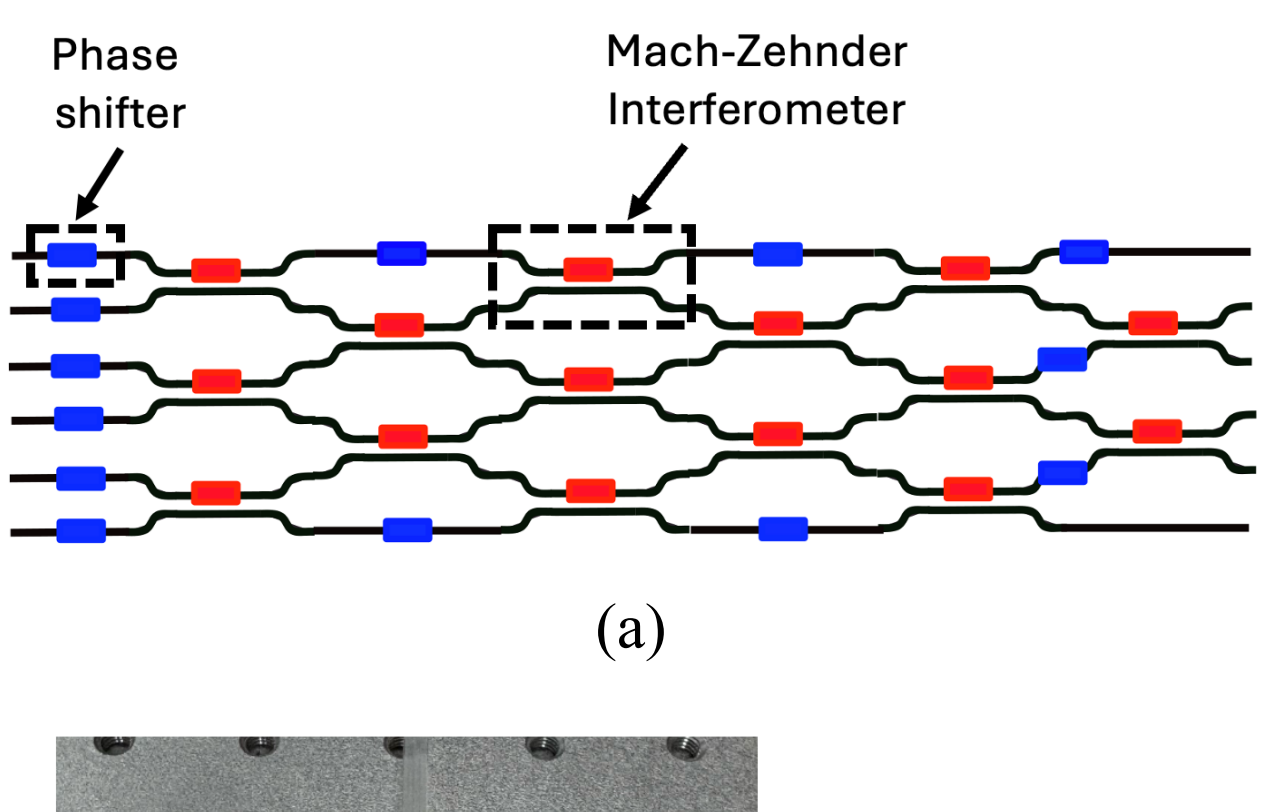

(a)

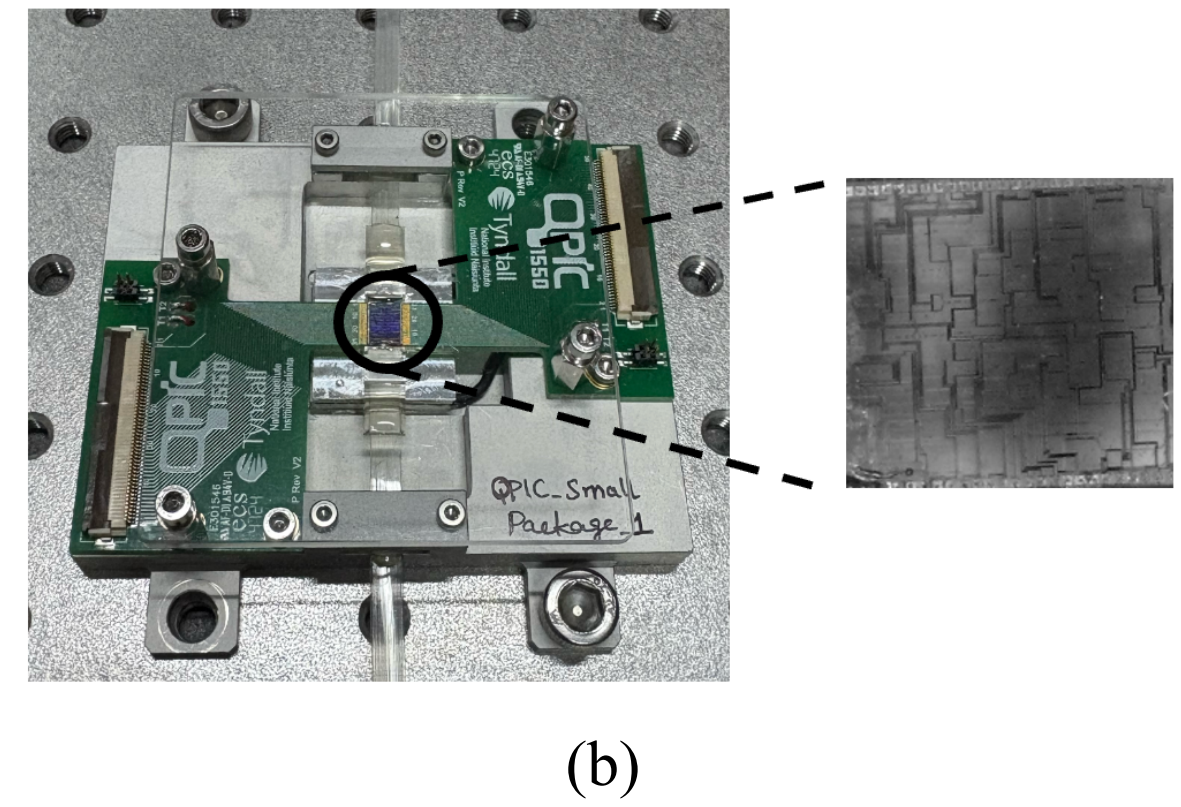

(b)

Fig. 1: (a) Schematic of the unitary MZI-based mesh. (b) The experimental packaged device (inset shows an image of the MZI array).

## II. Classical Silicon Photonics Physical Unclonable Function

### *A. Characterization of the SiN Photonic Rectangular Mesh*

The basis of the SP-PUF is an $6 \times 6$ rectangular integrated photonics mesh [22]. The mesh consists of 15 balanced Mach Zehnder Interferometers (MZIs) and 13 external phase shifters (Fig. 1). The phase difference at each MZI and external phase shifter is controlled by thermo-optic phase shifters (TOPs) of length equal to 750 $\mu m$. The total propagation losses are approximately equal to 11 dB and include grating coupler losses, propagation losses and insertion losses from MZIs. Rectangular arrays are reciprocal and linear systems that map input optical modes to output optical modes via a transformation $U \in \mathcal{C}^{6\times 6}$, that is approximately unitary up to uniform propagation and coupling losses (see Appendix A) [23]. The unitary matrix is fully defined by the phase vector $\phi \in \mathcal{R}^{28}$, which is controlled by the applied voltages $V \in \mathcal{R}^{28}$ through the thermo-optic relation described in Appendix A.

Collectively all passive offsets consitute the hardware fingerptint, originating from fabrication variations. The distribution of passive offsets can be extracted experimentally, by examining the nullification voltages $V_{null} \in \mathcal{R}^{28}$, defined by the condition $\phi\left(V_{null}\right) = \pi\, 1_{28}$. Using the voltage-to-phase mapping of Appendix A, the offsets are obtained as $\phi_{off} = \pi(1 - V_{null}^2/V_\pi^2)$ , where $V_\pi$ is the nominal π-voltage. When the TOP of a MZI is set at its nullification point, the output power at one of its port reaches its minimum value. A subset $V_{null}^{Sub} \in \mathcal{R}^{15}$ is estimated, for the voltages that control the MZI phase differences, by monitoring the output power under voltage sweeping. Thus, a transfer function is obtained per MZI as shown in Fig.2-a. By these transfer functions it is found that the voltage $V_{null}$ is different for each MZI, which is a direct consequence of the underlying phase uncertainty. The histogram with the distribution of the $V_{null}$ values among all the MZIs is depicted in Fig.2-b, indicating an approximately uniform distribution ranging from 7V to 8.6 V.

Before determining that sidewall roughess variability is the source of phase uncertainty, other physical scrambling mechanisms need to be excluded, with the most significant being the varying resistivities across TOPs and thermal crosstalk. For TOP resistivities, the distribution of heating power across the TOPs when a constant voltage of 1 V was applied was analyzed, providing an indirect measure of the ohmic resistance variability. The measured standard deviation was 0.0012 W. Given that the thermo-optic effect establishes an approximately linear relationship between temperature and optical phase, this limited variation in heating power suggests that resistance-induced phase fluctuations are relatively small.

Next, we investigate the effect of thermal variations due to crosstalk and global drifts. To mitigate global environmental thermal drifts, a thermo-electric closed-loop control system is employed to stabilize the chip temperature around 22°C, with measured variations on the order of 0.1°C at the package level. However, despite this global stabilization, local temperature gradients induced by the simultaneous operation of multiple TOPs can lead to additional phase fluctuations, which are not fully captured by the global temperature control. To this end we experimentally actuated one heater to its maximum nominal value of 12 V, which resulted in a power variation of less than 0.2 dB at the nearest passive MZI. These observations suggest that the impact of thermal crosstalk is limited.

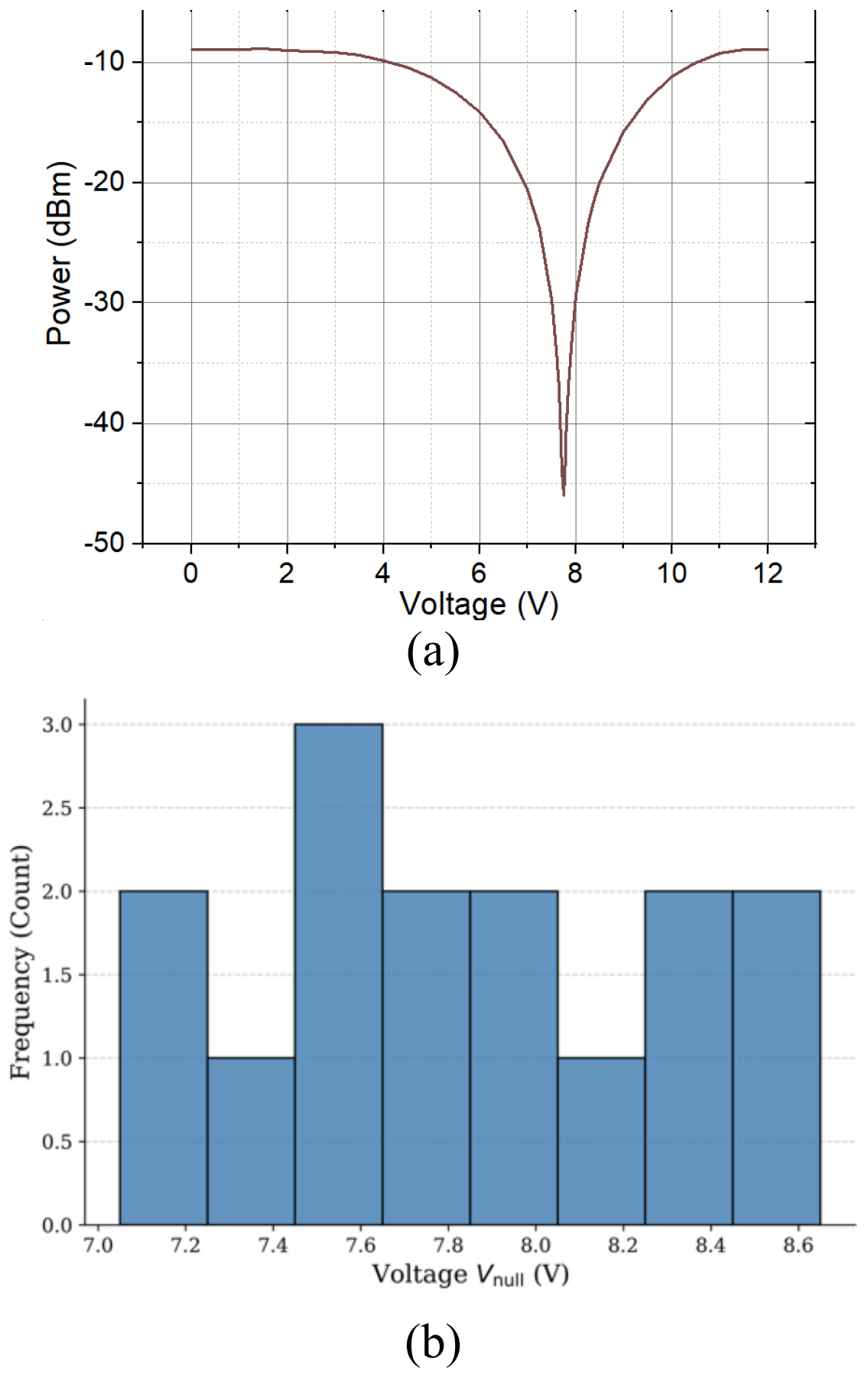

Fig. 2: (a) Power output of a MZI with respect to applied voltage difference. (b) Uniform distribution of the $V_\pi$ voltage among the 15 MZI of the experimental mesh.

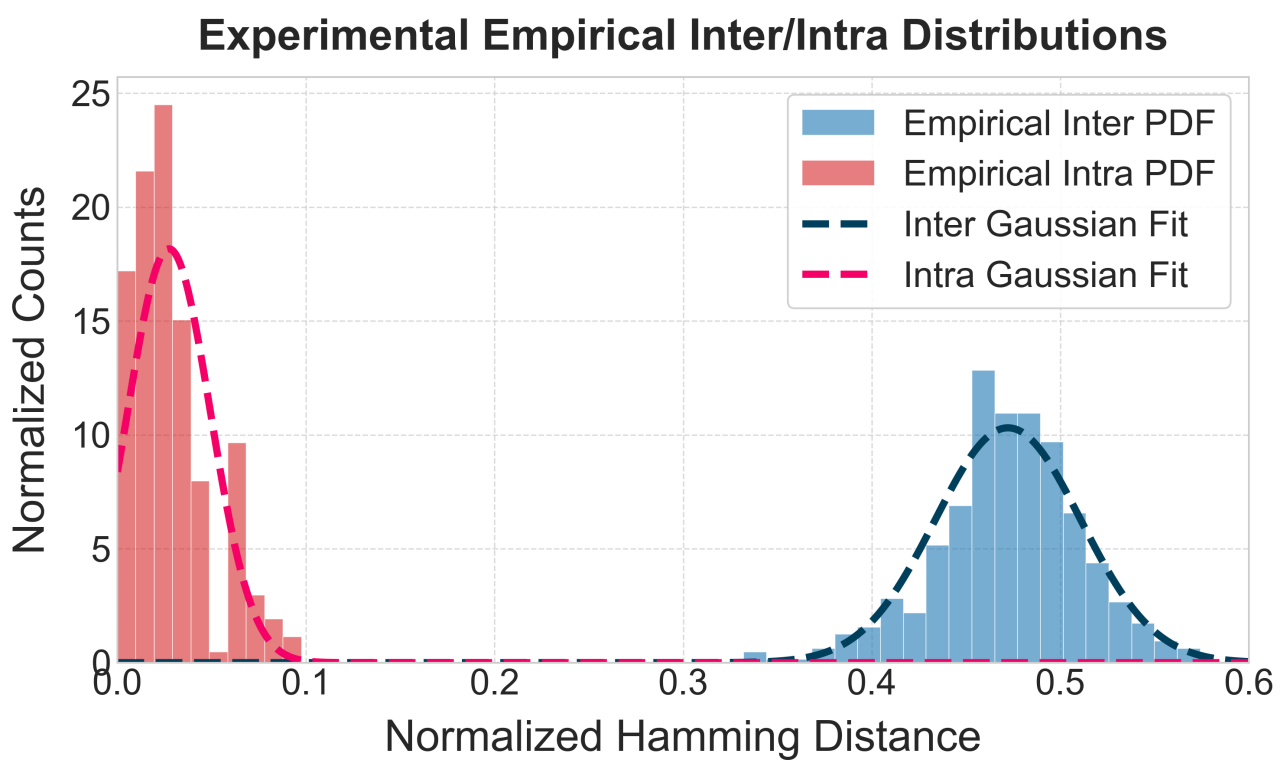

Fig. 3: Normalized Inter-Hamming and Intra-Hamming distributions for the strong SP-PUF.

To examine the effect of sidewall roughness variations, we use the measured $V_{null}$ distribution and the relation $\phi_{off} = \pi(1 - V_{null}^2/V_\pi^2)$, $V_\pi = \mathbb{E}[`V_{null}] = 7.5$ V. We infer a uniform distribution of static phase offsets with mean value $\sim 0.03$ rad, corresponding to an average effective-index variation on the order of $10^{-5}$ over a 750 $\mu m$ TOP length at 1550 nm. In addition, previously reported AFM mesurementes of sidewall roughness (0.5–3 nm) and correlation length (50–150 nm), together with the sensitivity of the effective index to waveguide width variations, indicate effective refractive index fluctuations in the range $10^{-5} - 10^{-4}$ [24]. This range is consistent with the experimentally inferred value, validating that sidewall roughness variations is the dominant source of the unique hardware fingerprint.

### *B. Experimental Verification of the Classical SP-PUF with the SiN mesh*

After device characterization, the system is examined as a PUF. The main objective is to define the CRPs and calculate robustness and randomness-entropy. Under the SP-PUF paradigm, the challenge is defined by the applied voltage settings and the response is the post-processed and binarized optical power profile . A continuous wave single mode field is injected into one or more input ports to sample the output spatial power profile. The SP-PUF maps externally applied random challenges to random power responses according to its unique and secret seed $\phi_{off}$, thus providing a means for multi-response generation through a larger internal entropy source.

In the experiment, the presented 6x6 SiN rectangular mesh has been employed [25]. A sub-area of 5 MZIs and 5 external phase shifters is tuned by TOPs, which ammounts to 10 actuators. An input optical power of 5 dBm provided by a laser-source is driven to the first port of the mesh. The output power values $P_{out}^{(m)} \in \mathcal{R}^6$ are recorded in dBm for $m = 1,2,\dots,30$ randomly chosen challenge vectors $V^{(m)} \in \mathcal{R}^{10}$. These vectors are drawn independently from a uniform distribution $U(5\,V, 10\,V)$ around the nullification points. It should be stated that the purpose of this experiment is the validation of the statistical uniqueness of the responses under random challenges, rather than exhaustively sampling the whole challenge space, as is standard practice in the experimental evaluation of strong PUFs [8], [9]. The outputs are post-processed to generate 256-bit keywords. First, the output power vectors are projected to a space of higher dimensionality equal to 32 via a fixed random Gaussian projector $G \in \mathcal{R}^{6\times 32}$ [26]. The resulting vector is quantized to 8-bit precision per component, yielding a 256-bit representation. To mitigate the system bias due to the asymmetric path lengths and the use of a single input port, three random and fixed permutations are applied to the bit sequences [27]. The XOR operation is applied between the three permutated sequences, yielding the final 256-bit key $b^{(m)} \in \{0,1\}^{256}$.

To estimate the randomness of the strong PUF, the fractional hamming distance between all pairs of generated responses $b^{(i)}$ and $b^{(j)}$, $i \neq j$ is computed. The resulting distances are plotted in a histogram and fitted with a Gaussian probability density function (PDF), thus providing the inter-PDF, presented in Fig.3. The Gaussian is characterized by a mean value of 46.25 % and standard deviation $\pm 3.5$ % indicating strong randomness.

To estimate the robustness of the system, 66 response keys of length 256 bits were recorded for a fixed challenge. The consecutive samples for robustness estimation were acquired within a 30 min time span, which are well above the characteristic thermal time constants of the photonic chip, thus offering a typical operational scenario. The fractional hamming distance between all the 66-bit sequences was calculated. These distances are plotted to a histogram and fitted by a Gaussian PDF, yielding the intra-PDF distribution. In Fig.3, the Gaussian is characterized by a mean value of 2.6 % and standard deviation $\pm 2.03$ %. This robustness metric can be mainly attributed to polarization induced fluctuations at the optical fiber interface that connects the laser to the photonic chip.

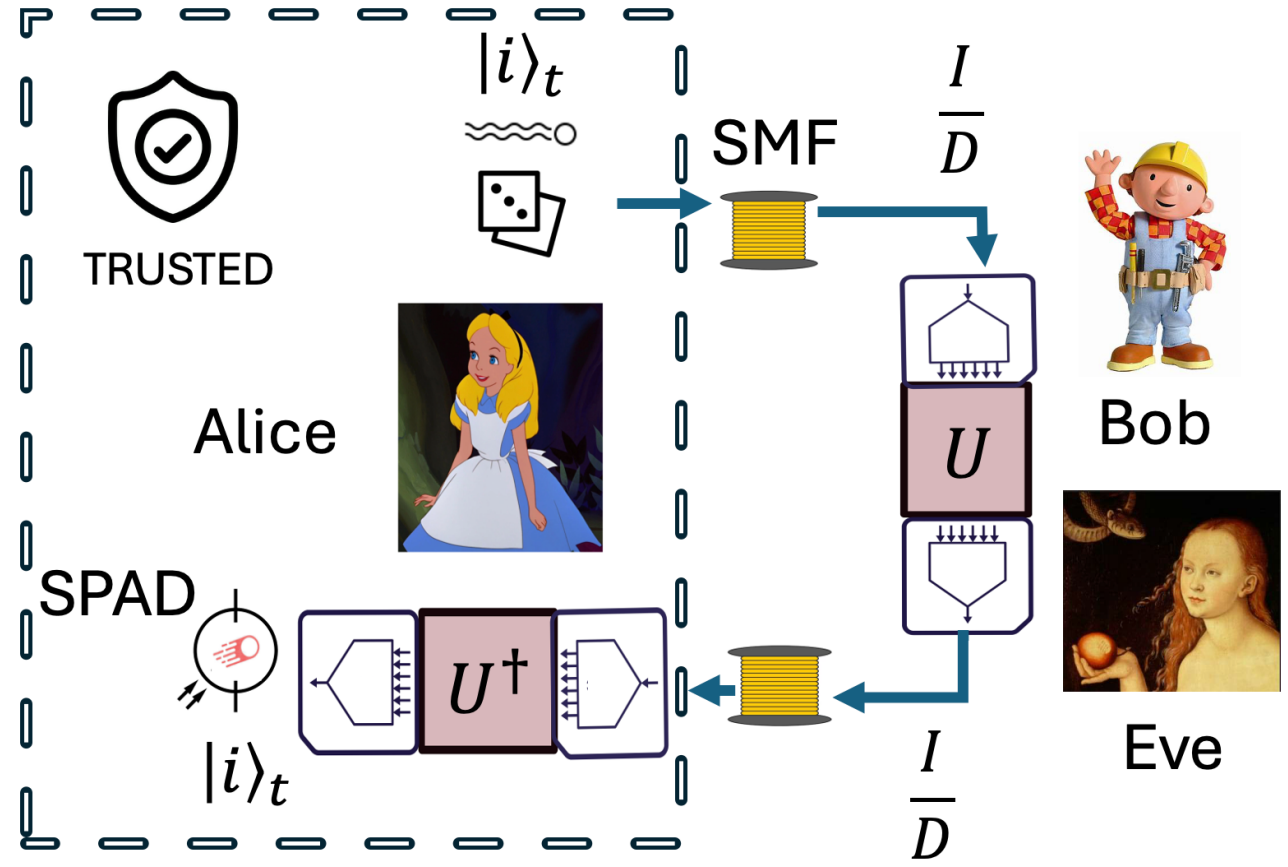

Fig. 4: The setup of the quantum-secure SP-PUF. Alice communicates with Bob via single photons (Fock states). The RNG produces MMS representations that reveal no information about the unitary transformation to Eve.

III. Quantum Physical Unclonable Function

The SP-PUF that was investigated experimentally in the previous section, is next considered as a physical authentication token operating under a quantum-secure protocol [28]. All numerical simulations presented in this section are based on experimentally extracted parameters from the fabricated SiN photonic mesh, ensuring realistic modelling conditions.

The setup for the QSP-PUF is presented in Fig. 4. In this concept, the authenticator (Alice) is regarded as a highly secure and trusted party, whereas the user (Bob) is regarded untrusted. In the enrollment phase, Alice characterizes the photonic chip (PUF), thus deriving knowledge about its static offsets $\phi_{off}$, before providing the device to the user (Bob).

There are three major attack scenarios, when a malevolent third-part (Eve) tries to be authenticated as Bob; (a) side channel attack, (b) cloning attack, (c) man-in-the-middle (MITM) attack. The first two cases are examined in this work, whereas the third type is shortly presetned in the Discussion section. In the side channel attack, the eavesdropper passively collects knowledge about the CRPs. For example, in the case of optical PUFs, this cold be fiber tapping. In the cloning attack, which involves an untrusted manufacturer, Eve has a PIC implemented on the same wafer, thus sharing fabrication conditions. She uses the mesh to authenticate herself as the legitimate user. Finally, in MITM attack, Eve transmits quantum-states and intercepts them from Bob's output during an authentication session, thus actively collecting knowledge about the device.

The QSP-PUF authentication protocol that aims to protect the user from these attacks operates as follows: An authentication session consists of multiple independent trials, equal to the number of photon shots provided by Alice to the channel. Per trial, Alice prepares a single photon $|1\rangle$, via spontaneous four wave mixing [29], spontanteous parametric down-conversion [30] or quantum-dots [31] and injects it to her 1 to D spatial routing network. The routing is controlled by a uniformly distributed random variable $i \in \{1,2,\ldots,D\}$ generated by a secure PRNG [32]. According to the outcome $i$, the router prepares the spatial-mode qudit state $|i\rangle = |0,0,\ldots,1,\ldots,0\rangle$, where the photon is in the $i$-th spatial mode. A trasncoder is used to map the qudit from the spatial to the temporal domain, by employing a set of delay lines and an optical switch [33]. This allows communication via a single mode fiber (SMF). Such transcoders are essential, as parallel fiber links would introduce uncontrolled phase differences, acting as a significant noise source. Since the transcoders preserve orthogonality, temporal and spatial encodings can be used interchangeably. The resulting temporal qudit is sent to Bob via the quantum channel.

Bob's device incorporates input and output transcoders for spatio-temporal mapping of quantum states [33], enabling both spatially implemented unitary transformations and SMF transmission. Bob applies his device-specific transformation $U_B = U_{PUF}$ producing the output pure state $U_B|i\rangle$, which is transmitted to Alice. At the receiver, Alice employs a similar PIC equipped with spatio-temporal transcoders. Owing to her knowledge of the passive offsets she reconstructs $U_{PUF}$ and implements its Hermitian conjugate $U_A = U_{PUF}^{\dagger}$ by reversing the propagation through the mesh. Consequently, the combined operation yields the pure state $U_A U_B|i\rangle = U_{PUF}^{\dagger} U_{PUF}|i\rangle = |i\rangle$. Since the final orthogonal mapping transforms the qudit from the spatial to the temporal domain, a single photon avalanche diode (SPAD) along with a tagger can be used. Because $i \in \{1,2,\ldots,D\}$, $D$ temporal windows are designated. Theoretically (ignoring losses and dark counts at the SPAD), when Bob wants to be authenticated through his mesh, each shot $i$ delivered by Alice will result to a click at the i-th time slot of the SPAD. In real conditions however, losses, enviromental drifts and dark counts will produce clicks at wrong time bins, directly affecting the system's stability. This concludes the case of a single trial.

A full authentication session consists of multiple trials, during which Alice transmits a sequence of single photons (Fock states) and records detection events (clicks). Only trials resulting in clicks are considered. The session ends when a predetermined number of clicks $N_{click}$ has been detected. Alice then maps the transmitted and received states and counts the number of incorrect detections (errors). A fuzzy threshold parameter $N_T$ is defined for decision making: if the number of errors satisfies $n \leq N_T$, the SP-PUF is accepted; otherwise, it is rejected. In this way, errors arising from stability limitations of the legitimate device are tolerated, whereas larger deviations introduced by an incorrect PIC lead to rejection.

### *A. Side Channel Atttack*

Protection from side channel attacks is based on the synergy between two quantum properties, the maximally mixed state (MMS) and the no-cloning theorem. In particular, due to the quantum nature of the system, the process as presented earlier is accessible only to Alice. Since she prepares the state, she has classical knowledge that the transmitted state is $|i\rangle$, without performing a measurement. In contrast, Eve (and Bob), who reside in the public channel, do not know which state has been sent prior to measurement. Because the index i is drawn uniformly, each state $|i\rangle, i = 1,\ldots,D$, is equally likely with probability $1/D$. From their perspective, the input is therefore described by a statistical mixture, represented by the density matrix [34]:

$$\rho_{in} = |\psi_{in}\rangle\langle\psi_{in}| = \frac{1}{D}\sum_{i=1}^{D}|i\rangle\langle i| = \frac{I}{D} \tag{2}$$

This differs from Alice's description, $\rho_{in} = |i\rangle\langle i|$, which reflects her knowledge of the prepared state. The discrepancy arises from information asymmetry prior to measurement. Since the states are uniformly sampled, the input state is described by an MMS from Eve's perspective. The MMS is inserted in Bob's unitary mesh and after propagation, the output ensemble remains maximally mixed, thus concealing the unitary transformation:

$$\rho_{out} = U_B\rho_{in}U_B^{\dagger} = U_B\frac{I}{D}U_B^{\dagger} = \frac{I}{D} \tag{3}$$

In contrast from Alice's viewpoint the output is $\rho_{out} = U_B|i\rangle\langle i|U_B^{\dagger}$ , which is a pure state. Physically, the MMS conceals the unitary transformation by removing any structure from the input ensemble, while the no-cloning theorem prevents Eve from copying quantum states. This is not the case in classical PUFs, where Eve by tampering can assess information for both classical I/O states at the same time, thus collecting CRPs that can be used to model or emulate the device. The presented analysis captures the fundamental physical mechanisms underlying security; a rigorous information-theoretic treatment is left for future work.

*B. Cloning Attack*

Eve may attempt an attack by employing a PIC fabricated using the same wafer process as the legitimate device. However, fabrication-induced variations, will produce a different unitary transformation $U_E \neq U_B$. Consequently, the final pure state after Alice's PIC is:

$$|\psi\rangle_{out} = U_A U_E|i\rangle \neq |i\rangle \tag{6}$$

Therefore, the final quantum state deviates from the initial state $|i\rangle$, which leads to mis-aligned clicks that exceed the dark-count noise level, rendering Eve detectable. It is therefore necessary to quantify how different $U_E$ is from $U_B$. Towards this direction, we compute the probability of an error per click $P(E|K) = 1 - P(C|K)$. Here, the probability $P(C|K)$ of a correct outcome (C) given a single detected click (K) is derived:

$$P(C|K) = \frac{(\eta\ tr(F)/D + d)(1-d)^{D-1}}{\eta + Dd} \tag{7}$$

The derivation appears in Appendix B. Here $F = |U_A U_B|^2$ , is a matrix containing all the probabilities of detecting a click at port $j$ given a photon sent at port $i$ and it is called the fidelity matrix. Moreover, $\eta$ is the probability for a photon to reach the receiver, which is dictated by the channel losses and the quantum efficiency of the SPAD. The dark count rate probability for the SPAD is denoted as $d$.

To quantify both the security of the scheme to cloning attack and its robustness to enviromental drifts, the false acceptance (FAR) and false rejection rates (FRR) will be defined [35]. The probability of observing less than $N_T$ errors among $N_{click}$ is given by the binomial cumulative distribution function (CDF):

$$P(n \leq N_T, N_{click}, P(E|K)) = \sum_{n=0}^{N_T}\binom{N_{click}}{n}P(E|K)^n(1 - P(E|K))^{N_{click}-n} \tag{8}$$

For the honest user (Bob) $P_B(E|K)$ is computed in terms of $F = |U_A U_B|^2$ , which refers the legitimate QSP-PUF. FRR is defined as the probability to reject the correct QSP-PUF, which means that more than $N_T$ (the fuzzy threshold determined by Alice) errors are observed for a session of $N_{click}$. This is expressed using the binomial CDF:

$$FRR = 1 - P\big(n \leq N_T, N_{click}, P_B(E|K)\big) \quad (9)$$

For the dishonest user (Eve) $P_E(E|K)$ is computed for $F = |U_A U_E|^2$. FAR is defined as the probability to accept the incorrect SP-PUF, which means that less than $N_T$ errors are observed for a session of $N_{click}$. This is also expressed using the binomial CDF:

$$FAR = P\big(n \leq N_T, N_{click}, P_E(E|K)\big) \quad (10)$$

Finally given a fuzzy threshold $N_T$, equal error rate (EER) is defined as the number of clicks $N^*_{click}$, where FAR is equal to FRR. At this operating point, security and robustness exhibit equal performance. The equal error rate (EER) serves as a key metric for PUF evaluation, with values below $10^{-6}$ indicating adequate performance and values below $10^{-9}$ indicating excellent performance [35].

To simulate the PIC, parameters extracted from Section II are used. The unitary transformation of the rectangular mesh is modeled according to Appendix A for D=6 ports, where $U = U(V, \phi_{off})$. Fabrication variations are incorporated by sampling the null voltages from a uniform distribution $V_{null} \sim \mathcal{U}(7, 8.6)\ V$ with a nominal value $V_\pi = 7.8$ V. Using the voltage-to-phase mapping defined in Appendix A and section II, these values determine the passive phase offsets $\phi_{off}$. All actuators are biased at $V = V_\pi 1$, such that the resulting phase vector is $\phi = \pi + \phi_{off}$, yielding the unitary transformation $U(\phi)$. A distinct device instance is generated by resampling $V_{null}$, which produces a different realization of the unitary matrix. Uniform propagation losses are assumed and set to 11 dB.

Environmental effects, such as polarization drifts, induce fluctuations in the measured output power. For single-port excitation, the output samples one column of $|U|^2$. By normalizing repeated measurements, these fluctuations translate into first-order probability variations with an estimated standard deviation of $\sigma \sim 2.4 \times 10^{-3}$. Assuming that these I/O-induced fluctuations are statistically similar across ports, the experimentally observed variations can be consistently extended to all columns. Therefore probabilities are sampled by a Gaussian with $P \sim \mathcal{N}(|U|^2, \sigma^2)$ .

The total scheme presented in Fig. 5, involves two PICs. The total loss budget for the PICs is equal to 22 dB. In order to include enviromental noise, the first-order probabilities are sampled by a Gaussian with $\hat{F} \sim \mathcal{N}(F, 2\sigma^2)$, where $F = |U_A U_B|^2$ is the fidelity matrix. Variance is multiplied by 2 to account for both PICs. For the SPAD, a dak count rate of $d = 10^{-6}$ is assumed based on 1000 counts/sec and a gating window of 1ns. Its efficiency is set equal to 25% [38].

First, the FRR is considered as a metric of robustness. Multi-click simulations are performed with 5000 Monte Carlo trials. Before the analysis, a unitary $U_s$ is sampled to implement Bob's transformation $U_B = U_s$. Alice's transformation is set as its Hermitial conjugate $U_A = U_S^\dagger$, known from the enrollment stage. The ideal fidelity matrix $F = |U_A U_B|^2$ is calculated and remains fixed for all trials. Afterwards, the analysis is initiated. For each trial a matrix $\hat{F}_B$ is sampled, yielding a different probability of error $P(E, \hat{F}_B|K)$ , which is used to evaluate the FRR for different $N_{click}$, $N_T$ parameters. Averaged results are presented in Fig.5-a. As $N_{click}$ is increased, the FRR is also increased because there is a higher probability for errors among a higher number of clicks. Increasing the accepted number of errors improves significantly the overall FRR as more miscounts are compensated.

FAR is calculated as a metric of the system's security when PICs from the same wafer are used by the attacker for authentication. Each PIC has a variance in phase offsets dictated by the distribution of nullification voltages. A Monte Carlo analysis of 5000 trials is performed to determine FAR. For each trial, a unitary mapping $U_E$ is sampled. Along with a matrix $U_A = U_S^\dagger$, that remains fixed during all trials, a fidelity matrix $\hat{F}_E$ is sampled, with mean value $F = |U_A U_E|^2$. Therefore, at each trial the probability of error $P(E, \hat{F}_E|K)$ takes different values and it is used to calculate the FAR. Averaged results are shown in Fig.5-b. From the graph as the tolerance $N_T$ is increased, the FAR is also increased, because with more acceptable errors there is a higher probability to mistakenly accept the dishonest user. However, increasing the number of clicks reduces exponentially the FAR, therefore reaching values lower than $10^{-45}$.

As in classical PUF literature, the equal error rate (EER) can be calculated, where robustness and security get equal (FAR=FRR) [35]. To estimate the EER a Monte Carlo analysis of 5000 trials is again followed. For each trial, two matrices $F_E, F_B$ are sampled as previously. For each threshold number $N_T$, we calculate the number of clicks $N^*_{click}$ that minimizes $|FAR - FRR|$. Click numbers ranging from 1 to 500 are examined. For each $N_T$, the EER is:

$$EER = \max\begin{pmatrix} FAR(N_T, N^*_{click}, F_E) \\ FRR(N_T, N^*_{click}, F_B) \end{pmatrix} \quad (11)$$

The averaged EERs are presented in Fig. 5(c). As the acceptance threshold $N_T$ increases, the EER decreases, indicating improved discrimination between legitimate and illegitimate devices. However, this improvement is achieved at the cost of requiring a larger number of detected clicks $N^*_{click}$. Consequently, higher values of $N_T$ lead to increased authentication latency, since more optical shots are needed to accumulate sufficient detection events under realistic loss conditions.

Fig. 5.a-c collectively reveal the fundamental trade-off between robustness and security in the proposed scheme. Given a threshold $N_T$, when $N_{cllick} < N^*_{click}$, the system operates in a regime where robustness dominates, meaning that legitimate users are reliably accepted but the tolerance to adversarial attempts is reduced. Conversely, for $N_{click} > N^*_{click}$, the system transitions to a security-dominated regime, where the probability of accepting an unauthorized device is significantly suppressed, at the expense of increased sensitivity to noise and imperfections. For example, for $N_T = 6$, the crossover point is $N^*_{click} = 28$, corresponding to an EER of approximately $10^{-6}$. Increasing the number of clicks to $N_{click} = 100 > N^*_{click}$ significantly reduces the FAR to $10^{-15}$, while the FRR increases to $\sim 10^{-3}$. This corresponds to approximately one false rejection per $10^3$ authentication sessions, in exchange for rendering cloning attacks practically infeasible.

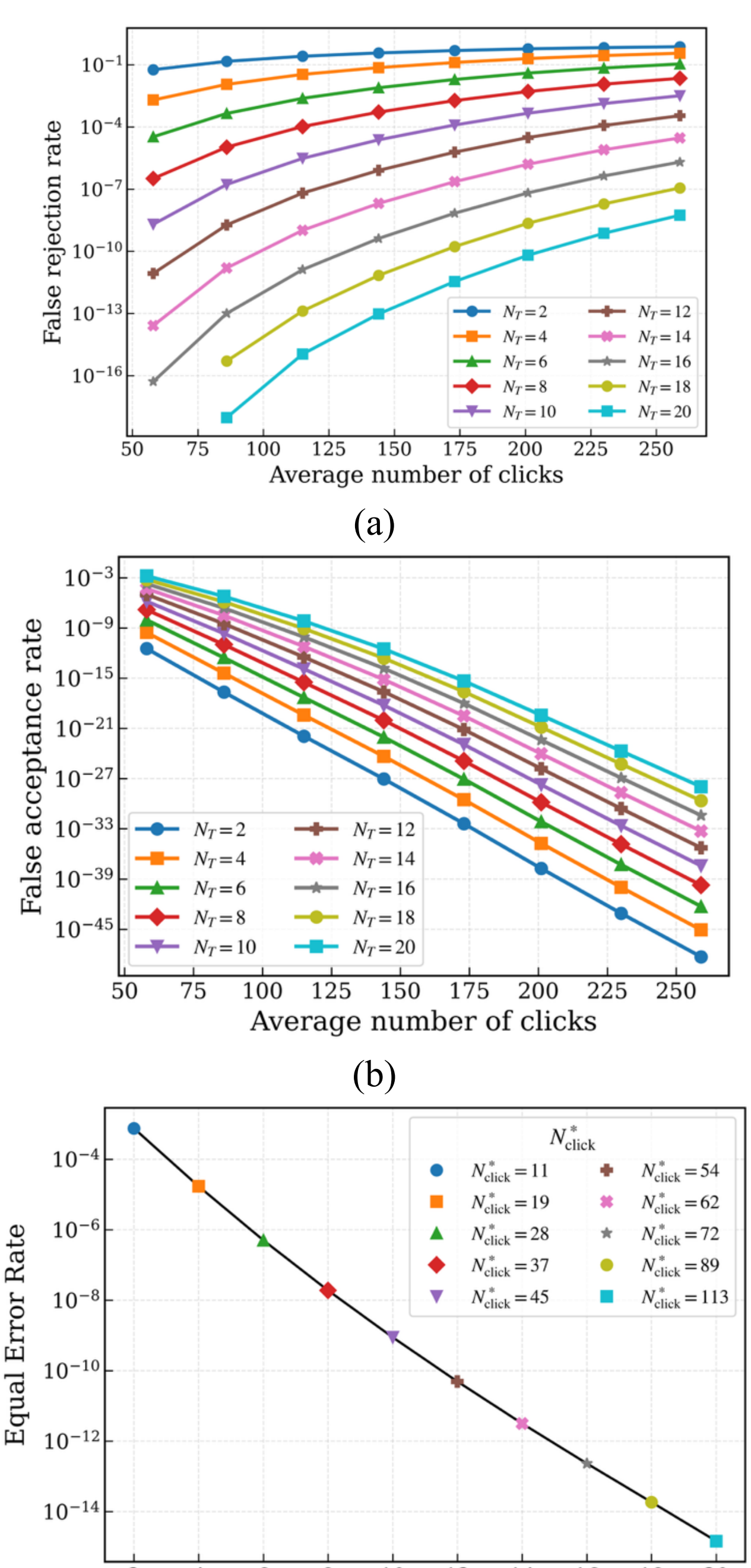

Fig. 5: (a) FAR and (b) FRR as a function of the number of detected clicks and of the error correction threshold for $D = 6$ ports. Fabrication variations for the FAR are modelled through the distribution of voltages $V_{null} \sim U(7, 8.6)V$. (c) EER as a function of threshold $N_T$ with the $N^*_{click}$ (legend), where FAR, FRR become equal.

## IV. Discussion

In the presented discussion, the first threat model is the side-channel attack, which assumes a passive eavesdropper that collects input and output quantum states. In this case, the unitary transformation is fully concealed by the MMS process. A more intrusive strategy would be a MITM attack, where Eve cuts momentarily the connection between Alice and Bob and sends her own states to probe the device. For a full characterization of a $D \times D$ unitary matrix, Eve must determine $D^2 - 1$ real parameters under different probe/measurement configurations [36]. For $D = 6$, this amounts to 35 real parameters. Under a session with $N_T = 16$ corrected errors, Alice waits for 72 clicks for the EER condition. In the case of a MITM attack, where Eve controls the channel for the whole authentication duration, this would yield $72/35 \approx 2$ detected events per parameter. To estimate a probability $p_i$ with precision $\sigma$, necessitates $N \sim p_i(1-p_i)/\sigma^2$ trials [37]. Using $E[p(1-p)] = E[p] - E[p^2] = 2^{-1} - 3^{-1} = 6^{-1}$ as a crude average, the precision achieved for $N = 2$ clicks is $\sigma \approx \sqrt{E[p(1-p)]/N} = 0.29$, meaning that the estimated matrix $F_E$ for the extracted $U_E$ exhibits large statistical fluctuations compared to $F_B$. Therefore, even by using all available shots in a communication session, Eve cannot deconstruct the unitary matrix with adequate precision. Moreover, Alice will detect Eve by monitoring the abnormally increased losses in the channel. Therefore, single photon probing is impractical. After each session, the classical challenge can be updated (set of phase shifters), thus resulting in a different unitary mapping, as experimentally validated in Section II.B, forcing Eve to re-start the characterization process. Other attacks, like probing with coherent states, carry elevated optical power that can be detected using standard watchdog detectors at Bob's input [38]. Additional key measures to protect the SP-PUF would be narrow-band filtering and optical isolation [38].

## IV. Conclusion

In this work, we provided experimental evidence of a SiN SP-PUF as a PUF, which was leveraged and analyzed numerically as the basis of a quantum secure authentication protocol. The PUF provides randomness at the scale of 45 % and robustness equal to 2 %. By employing quantum sources and detectors instead, when a single photon is driven to a random input port according to a uniform distribution, the state at the channel appears maximally mixed, concealing information about the fingerprint dictated transformation. The effect of fabrication variations is examined by calculating the false acceptance and the false rejection rates. The verifier by choosing the optimum number of detected clicks to terminate the authentication session and the number of accepted errors, can tune the error rates to become equal and as low as $10^{-14}$ exceptional security to the legitimate user. Therefore, quantum secure SP-PUFs appear as a promising platform for highly secure authentication applications, by combining the uncontrollable fabrication errors with foundational quantum security features.

## Appendix A

The basic block of a photonic unitary matrix of dimensionality $D$ is a physical node that consists of a MZI with an external phase shifter. For node with input ports $(k, k+1)$ at the $l$ −th layer, external phase $\gamma_{kl}[0,2\pi)$ and phase difference $\theta_{kl} \in [0,\pi]$, the transfer function is $U_{kl} \in \mathcal{C}^{D\times D}$ is:

$$U_{kl} = \begin{bmatrix} 1 & \cdots & \cdots & \cdots & 0 & 0 \\ 0 & \ddots & & & & \vdots \\ \vdots & & e^{i\gamma_{kl}}\,\mathrm{s}\,\theta_{kl} & e^{i\gamma l}\,\mathrm{s}\,\theta_{kl} & & \vdots \\ \vdots & & \mathrm{c}\,\theta_{kl} & -\,\mathrm{s}\,\theta_{kl} & & \vdots \\ \vdots & & & & \ddots & 0 \\ 0 & 0 & \cdots & \cdots & 0 & 1 \end{bmatrix} \quad \text{(A.1)}$$

Here $s(\cdot), c(\cdot)$ stand for sin and cos. By setting $D(D-1)/2$ such nodes according to the rectangular geometry presented in Fig.1, a unitary matrix $U \in \mathcal{C}^{D\times D}$ can be designed as:

$$U = \prod_{l=1}^{D} \prod_{k\in S_{l,D}} U_k(\theta_{kl}, \gamma_{kl})\ D(\gamma_1, \dots, \gamma_D) \quad \text{(A.2)}$$

Here, the layerwise product is left-multiplied from $l = D$ to 1. The second product implies the nodes per column, which acts on the modes $S_{l,D} = \{k \in [1,2,\dots,D \equiv l(mod\ 2)\}$. The matrix $D(\gamma_1, \dots, \gamma_D)$ is a diagonal matrix with every non-zero term being an offset $\exp(i\gamma_k), k = 1, \dots, D$. This layer is responsible for tuning the global phase of the output optical mode vector.

For notational simplicity all phase settings $\gamma_{kl}, \theta_{kl}, \gamma_k$ are collected in one vector $\phi \in \mathcal{R}^{N_\phi}, N_\phi = D^2$. In the experimental implementation, some phase shifters are omitted, resulting in $N_\phi < D^2$. The unitary matrix is a function of the phase vector $U(\phi)$.

The phases are controlled by voltages applied at thermo-optic actuators and the voltage to phase mapping is:

$$\phi = \phi_{off} + \pi \frac{V^2}{V_\pi^2} \tag{A.3}$$

Here, $V \in \mathcal{R}^{N_\phi}$ is the controllable vector that defines the phases. $\phi_{off} \in \mathcal{R}^{N_\phi}$ are the passive phase offsets -for voltages set at zero- that constitute the device fingerprint. $V_\pi$ is a nominal value shared by all phase shifters. Variations in $V_\pi$ across heaters are absorbed into $\phi_{off}$ for which the phase is equal to $\pi$ when no phase offsets is present. Therefore, the unitary matrix is a function of the voltages and passive offsets $U(V, \phi_{off})$.

## Appendix B

To compute eq. (7) of a correct measurement given a click, the Bayes rule is used as:

$$P(C|K) = \frac{P(K|C)P(C)}{P(K)} \tag{B.1}$$

Here, the events are defined as C: "a click occurs at the correct time-bin" and K:" a click occurs". Since a correct click necessarily implies that a click has occurred, it follows that $P(K|C) = 1$.

For an input qudit $|i\rangle$, a correct click arises if: $(C_1)$ the photon is detected at the i-th time-bin, or $(C_2)$ a dark count occurs at the same bin, while $(C_3)$ no spurious clicks occur at the remaining $D-1$ bins. These contributions are expressed as: $P(C_1) = \eta F(i,i) = \eta P(i|i)$, where $F$ contains the conditional probabilities, $P(C_2) = d$ and $P(C_3) = (1-d)^{D-1}$. Thus: $P(C|i) = (\eta F(i,i) + d)(1-d)^{D-1}$.

Averaging over uniformly distributed inputs, $P(i) = 1/D$, yields:

$$P(C) = \sum_{i=1}^{D} P(C|i)P(i) = \left[\frac{1}{D}\sum_{i=1}^{D} \eta F(i,i) + d\right](1-d)^{D-1} \tag{B.2}$$

The probability of a click is complementary to the no-click event, which occurs when no photon is detected and no dark count is registered, i.e.,$(1-\eta)(1-d)^D$. Therefore:

$$P(K) = 1 - (1-\eta)(1-d)^D \approx \eta + Dd \tag{B.3}$$

The approximation is satisfied when $\eta, d \ll 1$. Combining A1, A2, A3 and using $P(K|C) = 1$ , yields eq. (7):

$$P(C|K) = \frac{(\eta\ tr(F)/D + d)(1-d)^{D-1}}{\eta + Dd} \tag{B.4}$$